\documentclass[12pt,reprint,aps,groupedaddress,showpacs,nofootinbib,prd,twocolumn]{revtex4-2}

\usepackage{amsmath,amssymb,graphicx,amsfonts}
\usepackage{mathrsfs}
\usepackage{comment}
\usepackage{xcolor}
\usepackage[shortlabels]{enumitem}
\usepackage{hyperref}
\usepackage{url}
\usepackage{subfigure}

\begin{document}
\title{LitePIG: A Lite Parameter Inference system for the Gravitational wave in the millihertz band}

\author{Renjie Wang$^{1,2}$}
\author{Bin Hu$^{1,2}$}\thanks{corresponding author: bhu@bnu.edu.cn}
\affiliation{$^{1}$Institute for Frontier in Astronomy and Astrophysics, Beijing Normal University, Beijing, 102206, China}
\affiliation{$^{2}$Department of Astronomy, Beijing Normal University, Beijing, 100875, China}

\begin{abstract}
We present a python based parameter inference system for the gravitational wave (GW) measured in the millihertz band. 
This system includes the following features: the GW waveform originated from the massive black hole binaries (MBHB), the stationary instrumental gaussian noise, the higher-order harmonic modes, the full response function from the time delay interferometry (TDI) and the gaussian likelihood function with the dynamic nested parameter sampler. 
In particular, we highlight the role of higher-order modes. 
By including these modes, the luminosity distance estimation precision can be improved roughly by a factor of 50, compared with the case with only the leading order ($\ell=2,|m|=2$) mode. This is due to the response function of different harmonic modes on the inclination angle are different. Hence, it can help to break the distance-inclination degeneracy. Furthermore, we show the robustness of testing general relativity (GR) by using the higher-order harmonics. Our results show that the GW from MBHB can simultaneously constrain four of the higher harmonic amplitudes (deviation from GR) with a precision of 
$c_{21}=0.54^{+0.61}_{-0.82}$, $c_{32}=-0.65^{+0.22}_{-0.08}$, $c_{33}=0.56^{+0.60}_{-0.76}$ and $c_{44}=1.57^{+2.34}_{-1.90}$, respectively. 
\end{abstract}
	
\maketitle

\section{Introduction}
Gravitational waves (GWs) from compact binary coalescence (CBC) events open a new observational window to explore the nature of the universe.
The LIGO \cite{LIGOScientific:2014pky}-Virgo \cite{VIRGO:2014yos}-KAGRA \cite{KAGRA:2018plz} (LVK) scientific collaboration has detected 90 GWs from CBCs up to the end of the third observing run(O3) \cite{LIGOScientific:2021djp}. 
These CBC events can be used to study the properties of stellar-mass black holes and neutron stars.
The future space-borne gravitational wave observatory, such as the Laser Interferometer Space Antenna (LISA) \cite{LISA:2017pwj}, Taiji \cite{Hu:2017mde} and TianQin \cite{Luo:2020bls} will be able to measure the GW events in the millihertz band, which contains many sources including massive black hole binaries (MBHB) \cite{Barausse:2020kjy}, extreme mass-ratio insprial(EMRI) \cite{Babak:2017tow} and white dwarf binaries \cite{Littenberg:2020bxy}, {\it etc}.
MBHBs are one of the main scientific goals of the space-borne gravitational wave observatory because of their high detection significance. Hence, these objects are suitable for the precise astrophysics and cosmology studies \cite{Schutz:1986gp,Zhu:2021bpp,Yang:2020yoc,Holz:2005df}. 

In general, GW signals are the superposition of multiple harmonic modes \cite{Thorne:1980ru}. 
The dominant harmonic is the $(\ell,|m|)=(2,2)$ harmonics.
Compared with the (2,2) mode, the higher-order harmonics are much weaker. 
Two important intrinsic parameters affecting amplitude of each higher-order modes are mass ratio and total mass \cite{Mills:2020thr}.
The spin parameter entering at higher post-Newtonian (PN) order, have an significant impact on the waveform \cite{Mishra:2016whh}. 
The higher-order harmonics can bring in new dependencies on the mass ratio, component spins, and inclination angle into the waveform\cite{Divyajyoti:2021uty,VanDenBroeck:2006ar,Arun:2008kb}.
The importance of including these higher-order harmonics into the waveform is that one can break the degeneracy between the luminosity distance and inclination angle\cite{Ohme:2013nsa,Usman:2018imj}. 
In recent observations of GW190412 \cite{LIGOScientific:2020stg} and GW190814 \cite{LIGOScientific:2020zkf}, which generated from CBCs with significantly asymmetric component masses, confirm the presence of higher-order harmonics emission at high confidence level.
The network with two 3G ground-based detectors are promising to detect the higher-order harmonics \cite{Divyajyoti:2021uty}.
Furthermore, including higher-order harmonics can increase the signal-to-noise ratio for Intermediate mass ratio inspiral (IMRI) binaries and improve the measurement of IMRI source properties \cite{Islam:2021zee}.
For an equal-mass neutron star merger system at distance similar to GW170817, the inclusion of higher-order modes leads to improvements of the distance and inclination and allow for percent-level measurements of Hubble parameter \cite{CalderonBustillo:2020kcg}.
Further, the higher-order harmonics can be used to test GR \cite{Mishra:2010tp,Puecher:2022sfm}. 

In our work, we present a python based parameter inference system for the gravitational wave (GW) measured in the millihertz band. In particular, we highlight the importance of including the higher-order harmonics in the parameter estimation. 
In section \ref{sec:lisa}, taking LISA as an example, we demonstrate the methodology of GW signal generation, the time delay interferometry response and the low frequency approximation, which is currently adopted.  
In section \ref{sec:inference}, we describe the parameter likelihood contruction as well as the sampling method. In section \ref{sec:valid}, we show the luminosity distance parameter estimation from MBHB as an example. Furthermore, we show the capability of testing general relativity (GR) with higher-order harmonics. We draw the conclusion at the end.

\section{LISA response and waveform generation}
\label{sec:lisa}

In this section, we will take LISA as an example to illustrate our methodology of the GW waveform generation. 

\subsection{The GW signal}
The GW waveform in the transverse-traceless gauge is described by the two polarizations $h_+$, $h_\times$.
Furthermore, $h_+$, $h_\times$ can be decomposed into the spherical harmonic modes, $h_{\ell m}$, by using the spin-weighted spherical harmonics as functions of the inclination angle $\iota$ and coalescence phase $\phi_c$. The waveform can be expressed as
\begin{equation}
    h_+ -i h_{\times} = \sum_{\ell\geq2}\sum_{m=-\ell}^{\ell}\ _{-2}Y_{\ell m}(\iota,\phi_c)h_{\ell m}\;. 
\label{decompose}
\end{equation}
The spin-weighted spherical harmonics for the modes $_{-2}Y_{lm}(\iota,\phi_c)$ can be found in Appendix A of \cite{Mills:2020thr}. The dominant harmonic is $h_{22}$, all the others are called higher-order modes.
Furthermore, one can translate the mode decomposition Eq.(\ref{decompose}) into the Fourier domain
\begin{equation}
    h_+ = \frac{1}{2}\sum_{l,m}(_{-2}Y_{lm}(\iota,\phi_c)h_{lm}+\ _{-2}Y_{lm}^*(\iota,\phi_c)h_{lm}^*)\;,
\end{equation}
\begin{equation}
    h_{\times} = \frac{i}{2}\sum_{l,m}(_{-2}Y_{lm}(\iota,\phi_c)h_{lm}-\ _{-2}Y_{lm}^*(\iota,\phi_c)h_{lm}^*)\;.
\end{equation}
For non-precessing binary system, an exact symmetry relation between modes allows us to write
\begin{equation}
    h_{l,-m}=(-1)^lh_{lm}^*\;.
\end{equation}
Therefore, we can further write
\begin{equation}
    h_{+,\times} = \sum_{l,m} K_{lm}^{+,\times}(\iota,\phi_c)h_{lm}\;,
\end{equation}
with
\begin{equation}
    K_{lm}^+(\iota,\phi_c) =\frac{1}{2}\big(_{-2}Y_{lm}(\iota,\phi_c)+(-1)^l\ _{-2}Y_{lm}^*(\iota,\phi_c)\big)\;,
\end{equation}
\begin{equation}
    K_{lm}^{\times}(\iota,\phi_c) =\frac{1}{2}\big(_{-2}Y_{lm}(\iota,\phi_c)-(-1)^l\ _{-2}Y_{lm}^*(\iota,\phi_c)\big)\;.
\end{equation}
In the Fourier domain, we often use the one side frequency spectrum, namely either keep the positive or the negative parts depending on the sign of $m$ \cite{Mills:2020thr,Marsat:2020rtl,Marsat:2018oam}. This approximation is valid in particular where the stationary phase approximation can be used. Hence, we assume
\begin{equation}
    \tilde{h}_{lm}(f)\simeq 0
    \begin{cases}
    f>0,m<0\\
    f<0,m>0\;,
    \end{cases}
\end{equation}
and neglect modes $h_{l0}$. With this approximation we can obtain explicit expressions for positive frequency modes
\begin{equation}
    \tilde{h}_{+,\times}(f) = \sum_{l}\sum_{m>0}K^{+,\times}_{lm}(\iota,\phi_c)\tilde{h}_{lm}(f)\;.
\end{equation}

\begin{figure}[h]
    \centering
    \subfigure{
    \begin{minipage}[t]{1.0\linewidth}
    \includegraphics[width=1.0\linewidth]{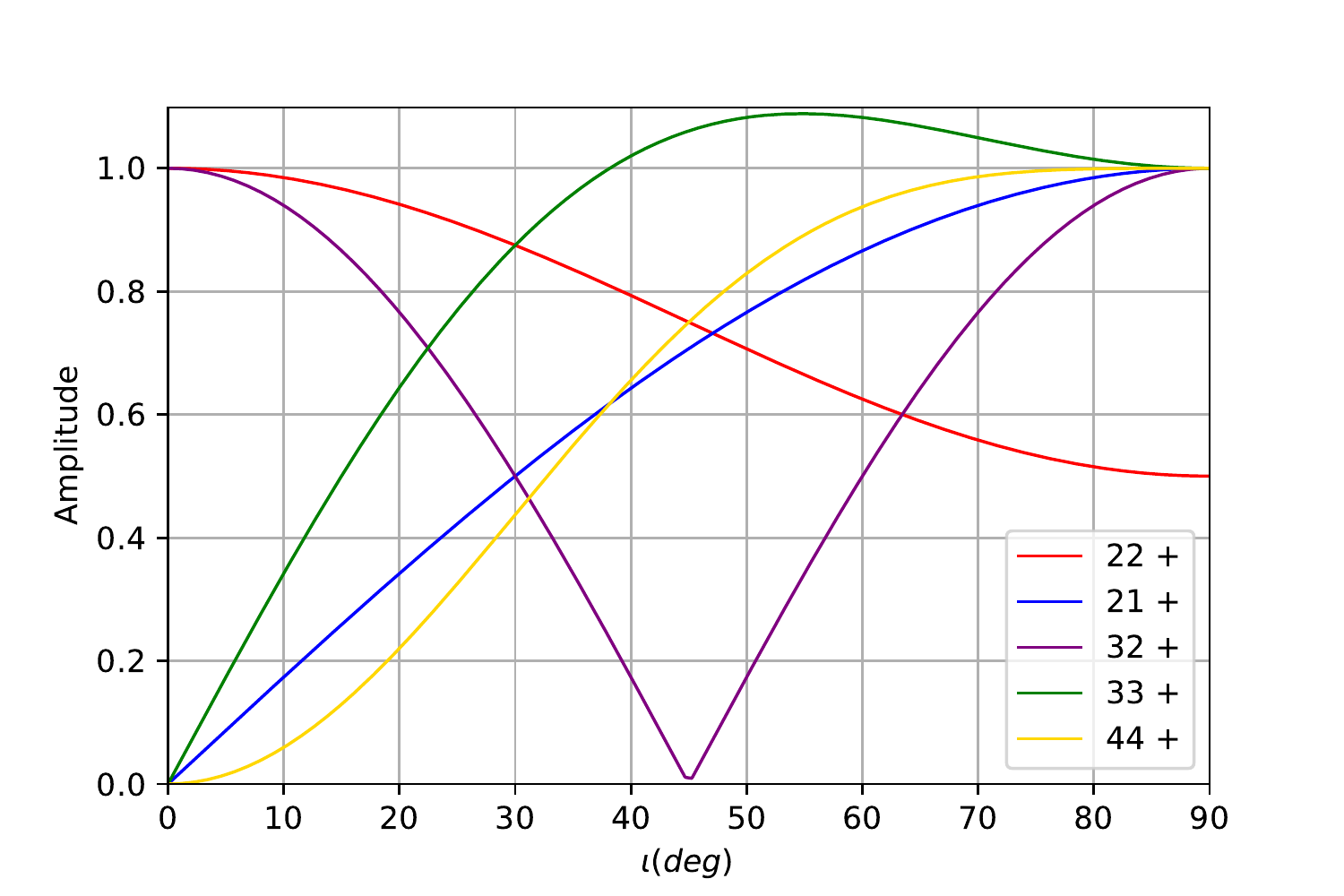}
    \end{minipage}
    }
    \subfigure{
    \begin{minipage}[t]{1.0\linewidth}
    \includegraphics[width=1.0\linewidth]{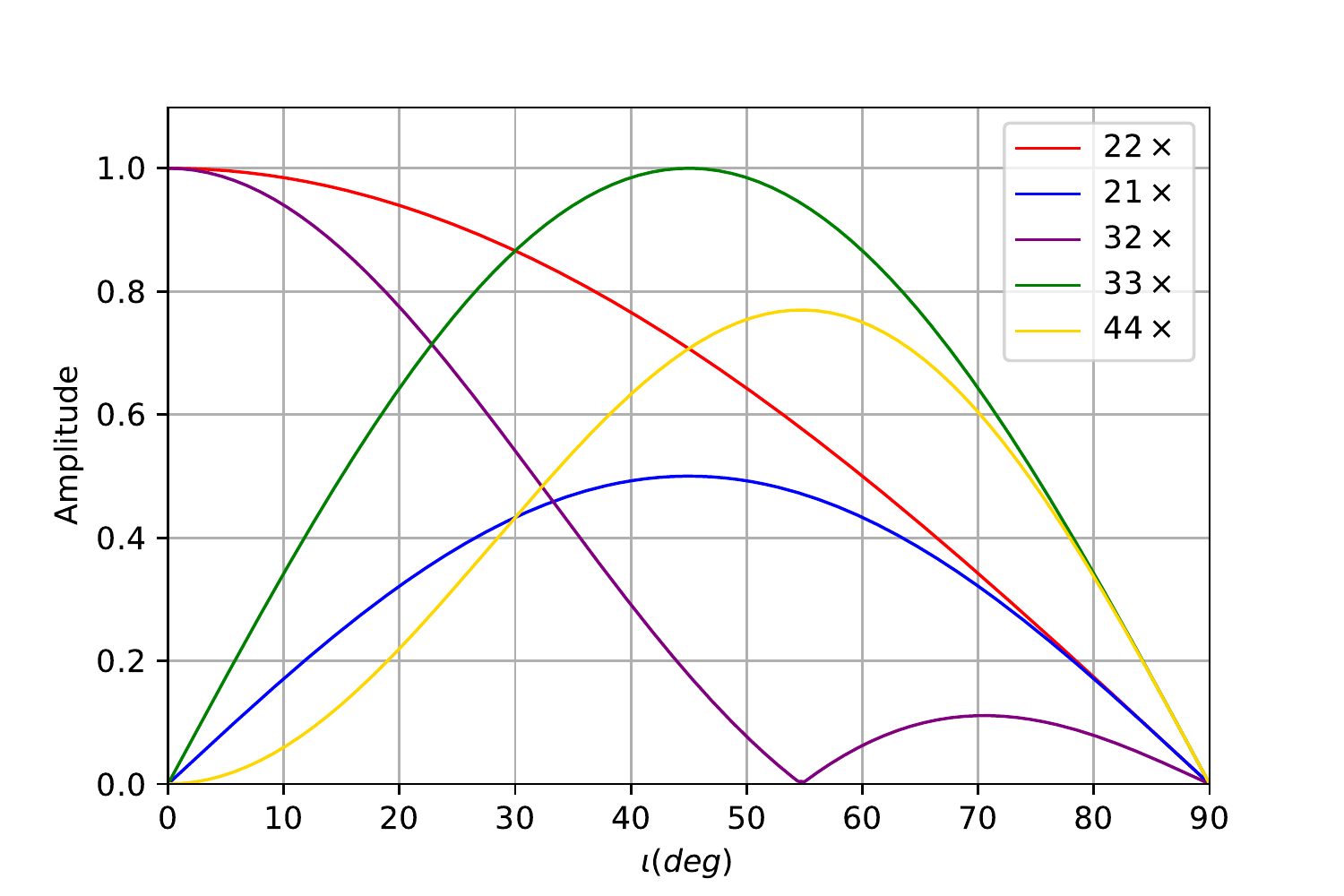}
    \end{minipage}
    }
\caption{The amplitude of each harmonics as a function of the inclination $\iota$. The red, blue, purple, green and gold lines denote the amplitude of some harmonics including the dominant modes $\ell=|m|=2$ and a set of higher-order harmonics, $(\ell,|m|)=(2,1),(3,3),(3,2),(4,4)$. The first row is result for the plus polarization while the second row is for the cross polarization.}
\label{fig:modes_amp}
\end{figure}

The GW waveform can be expressed as a superposition of individual harmonics. Fig (\ref{fig:modes_amp}) shows the dependence of the modes on inclination. For the (2,2) mode, the plus polarization peaks at face-on configuration, while (2,1),(3,3) and (4,4) modes vanish at face-on. The amplitude of (3,2) modes can reach maximum at both face-on and edge-on. Similarly, the amplitude of the cross polarization dependence on inclination angle are different for each modes.
Because of these differences, the inclination measurement can be improved when higher-order modes are measured \cite{Mills:2020thr}. Therefore, the well-known degeneracy between distance and inclination angle can be broken.

\subsection{Time Delay Interferometry (TDI): full response}
For the ground-based laser interferometers, such as LIGO  \cite{LIGOScientific:2014pky}, one can keep the same arm length up to the picometer level. Hence, the laser frequency noise can be cancelled very precisely. However, the spaceborne GW observatory will have much longer arm lengths. Compared to the ground-based interferometers, the GW space mission will be impossible to maintain equal arm lengths between spacecraft pairs. Because of the dynamics of mission orbit, the laser frequency noise cannot be cancelled to the level of measuring GW signal. To exactly cancel the laser frequency noise, time delay interferometry (TDI) technique is proposed \cite{1999ApJ...527..814A}. TDI had been well studied for the first-generation \cite{Tinto:2000qij,Dhurandhar:2001tct} and the the second-generation \cite{Vallisneri:2004bn}.

The TDI technique is used to construct a new set of observables from delayed combinations of $y_{sr}$ to cancel the laser frequency noise. 
Using the notation $y_{sr,nL}=y_{sr}(t-nL)$, the first generation TDI observable X is given by
\begin{equation}
\begin{split}
X= y_{31} +y_{13,L} + (y_{21}+y_{12,L})_{,2L}\\
-(y_{21}+y_{12,L})-(y_{31}+y_{13,L})_{,2L} \;.
\end{split}
\end{equation}
The other two observables Y and Z are obtained by cyclic permutation. The first generation TDI observables (X,Y,Z) are correlated in their noise properties. They can be transformed into uncorrelated observables (A,E,T). Thus, the second generation TDI observables A, E and T are expressed as
\begin{equation}
A=\frac{1}{\sqrt{2}}(Z-X)\;, \\
\end{equation}
\begin{equation}
E=\frac{1}{\sqrt{6}}(X-2Y+Z)\;, \\
\end{equation}
\begin{equation}
T=\frac{1}{\sqrt{3}}(X+Y+Z)\;. \\
\end{equation}
These channels are independent and uncorrelated.

The source frame waveforms can be represented as a combination of harmonics with the amplitude $A(f)$ and phase $\Psi(f)$
\begin{equation}
\tilde{h}_{lm}(f) = A_{lm}(f) e^{-i\Psi_{lm}(f)}\;.
\end{equation}
Currently, there are five harmonics included in the \texttt{IMRPhenomXHM} template \cite{Garcia-Quiros:2020qpx}, namely $(l,|m|)=(2,2),(2,1),(3,3),(3,2),(4,4)$. A complex transfer function $\mathcal{T}(f,t_{lm}(f))$ is used to transform the source-frame waveform into the TDI observabels \cite{Marsat:2018oam,Marsat:2020rtl}.
This function is determined by the extrinsic parameters $(\iota,\lambda,\beta,\psi,\phi_{0},t_{0})$. Because of the evolution of the LISA constellation, transfer function is temporal- and frequency-dependent.
The time-frequency dependence for each harmonics, $t_{lm}(f)$ is defined by the stationary phase approximation (SPA)
\begin{equation}
t_{lm}(f) = t_0 - \frac{1}{2\pi}\frac{d\Psi_{lm}(f)}{df}
\end{equation} 
The signal of each harmonics and for each TDI channel is given by
\begin{equation}
\tilde{h}^{A,E,T}_{lm}(f)= \mathcal{T}^{A,E,T}(f,t_{lm}(f))\tilde{h}_{lm}(f)
\end{equation}

\begin{figure}[h]
    \centering
    \includegraphics[width=1.0\linewidth]{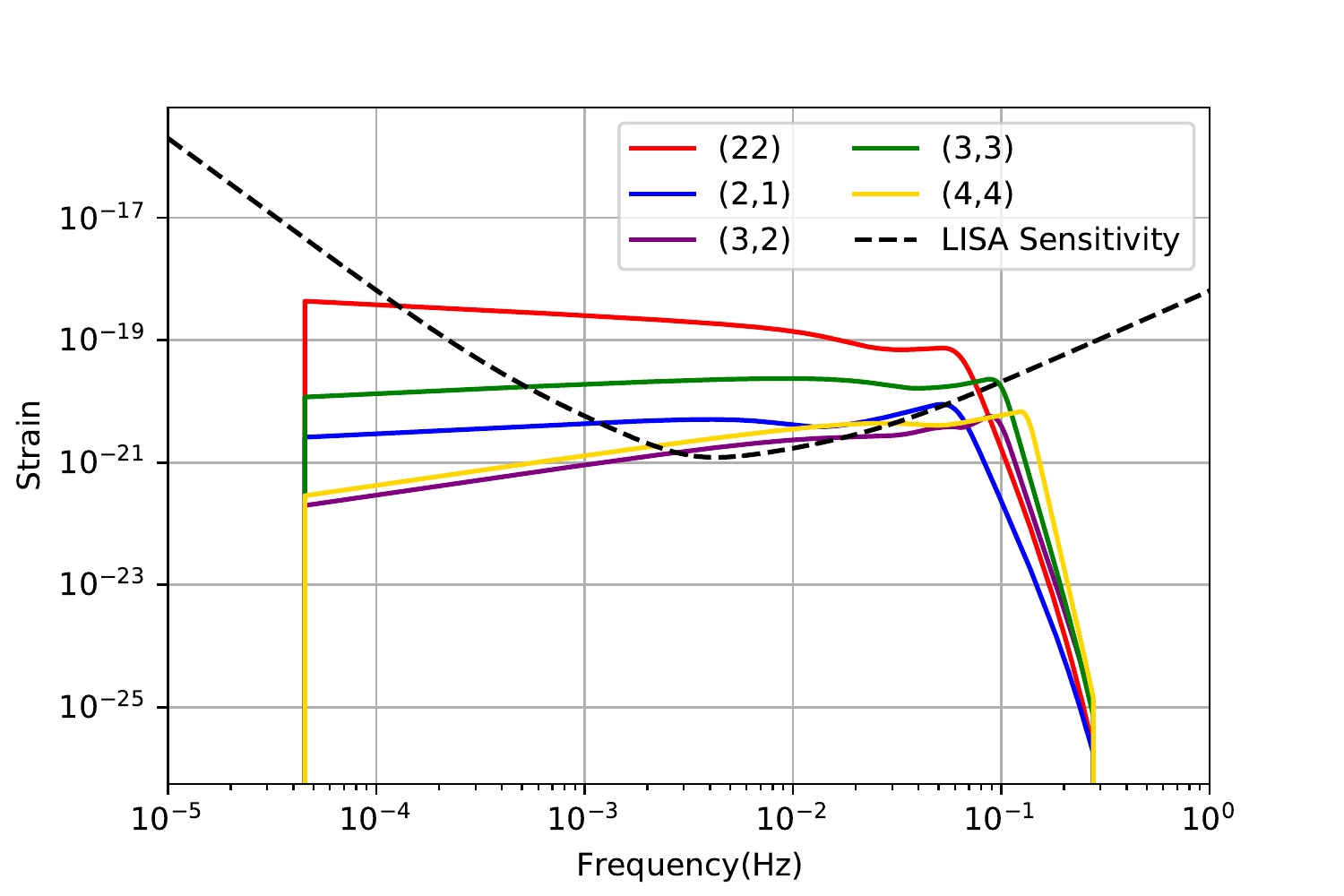}
    \caption{Characteristic strain of each harmonic mode. This displays the amplitude of each mode using the \texttt{IMRPhenomXHM} waveform template. Note that the plot displays the theoretical signal without convolving the response function. The black dotted line is LISA sensitivity curve from \cite{Robson:2018ifk}}.
    \label{fig:charac_strain}
\end{figure}

Fig.(\ref{fig:charac_strain}) shows the characteristic strain for each harmonic mode $\tilde{h}_{lm}$ in frequency domain. The (2,2) mode $\tilde{h}_{22}$ is the most dominant one and followed by (3,3) and (2,1), sequentially. These curves are the theoretical template, the observation response function has not yet been added. 

\begin{figure}[h]
    \centering
    \includegraphics[width=1.0\linewidth]{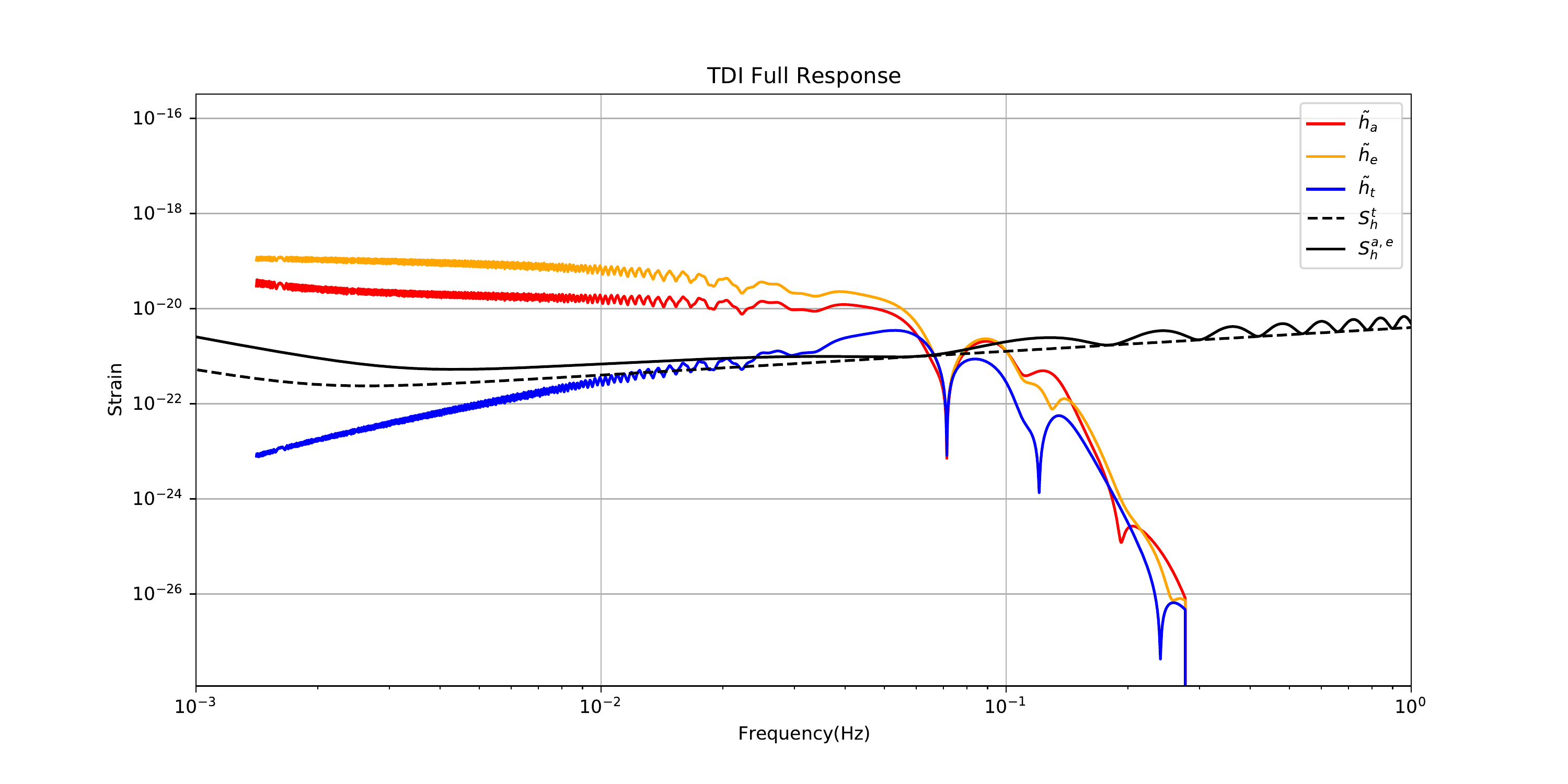}
    \caption{The characteristic strain in the three TDI channels. The red, orange and blue curves represent the characteristic strain of the signals defined by Eq. (\ref{tdi}) : $h^2_c=|f^2\tilde{h}_{a,e,t}(f)|$. The black solid and dotted curves show the characteristics strain with the reduced noise PSD $S^{a,e,t}_h(f)$ from Eq. (\ref{PSD}).}
    \label{fig:psd}
\end{figure}

To compensate the fast oscillatory in the high frequency range and avoid the numerical instability, we can rescale the TDI observables by prefactors which are common to both signal and noise \cite{Marsat:2020rtl}. Thus, the TDI channels in the frequency domain can be written as
\begin{equation}
\tilde{a},\tilde{e} = \frac{e^{-2i\pi fL}}{i\sqrt{2}{\rm sin}(2\pi fL)}\times \tilde{A},\tilde{E}\;,
\end{equation}
\begin{equation}
\tilde{t} = \frac{e^{-3i\pi fL}}{2\sqrt{2}{\rm sin}(\pi fL){\rm sin}(2\pi fL)}\times \tilde{T}\;.
\end{equation}
By combining $\tilde{h}_{lm}(f)$ and mode transfer function $\mathcal{T}(f,t_{lm}(f))$, we can produce the templates for the reduced channels
\begin{equation}
\tilde{a},\tilde{e},\tilde{t}= \sum_{l,m}\mathcal{T}^{lm}_{a,e,t}\tilde{h}_{lm}\;.
\end{equation}
Factoring out the same sine square function from the noise power spectral density (PSD) 
\begin{equation}
    S^{A}_{n},S^{E}_{n}=2{\rm sin}^2(2\pi fL)\times S^{a}_{n},S^{e}_n\;,
\end{equation}
\begin{equation}
    S^{T}_{n}=8{\rm sin}^2(\pi fL){\rm sin}^2(2\pi fL)\times S^{t}_{n}\;,
\end{equation}
the reduced PSD reads 
\begin{eqnarray}
S^{a}_{n} = S^{e}_{n} &=& 2(3+2{\rm cos}(2\pi fL)+{\rm cos}(4\pi fL))S^{pm}(f)\nonumber\\
&&+(2+{\rm cos}(2\pi fL))S^{op}(f)
\end{eqnarray}
\begin{equation}
S^{t}_{n} = 4{\rm sin}^{2}(2\pi fL)S^{pm}(f)+S^{op}(f)\;,
\end{equation}
where $S^{pm}$ is the test-mass noise PSD and $S^{op}$ is the optical noise PSD. The corresponding values for LISA can be found in Appendix A4 of \cite{Marsat:2020rtl}.
It is also useful to introduce the following notations 
\begin{equation}
\tilde{h}_{a,e,t} = \frac{1}{-6i\pi fL} \times \tilde{a},\tilde{e},\tilde{t}\;.
\label{tdi}
\end{equation}
Thus, we can define a noise PSD associated to the TDI observables (\ref{tdi}) as
\begin{equation}
S^{a,e,t}_{h}(f) = \frac{S^{a,e,t}_{n}(f)}{(6\pi fL)^2}\;.
\label{PSD}
\end{equation}
Moreover, the characteristic noise strain can be defined to be 
\begin{equation}
    S^{a,e,t}_c(f)= fS^{a,e,t}_{h}(f)\;. 
\label{nosie_strain}
\end{equation}
In Fig (\ref{fig:psd}), the solid curves show the characteristic strain (\ref{tdi}) and the reduced noise PSD (\ref{PSD}) for three TDI channels. One can see that the A channel is similar to the E channel, while the T channel is noise dominated. 

\subsection{The low-frequency limit}
Though we can apply the full response to perform parameter estimation, it will be useful to consider some limits to simplify calculations.
Here, we use the low-frequency approximation to the LISA response. 
When $f\ll f_L=1/L=0.12Hz$ \cite{Marsat:2020rtl}, the T-channel can be neglected. And in this low-frequency approximation, the response for the other two TDI observables in Eq. (\ref{tdi}) are given by 
\begin{equation}
    \tilde{h}_{a,e} = F^{+}_{a,e}(\lambda_L,\beta_L,\psi_L)\tilde{h}_{+} + F^{\times}_{a,e}(\lambda_L,\beta_L,\psi_L)\tilde{h}_{\times}\;,
\label{LFtdi}
\end{equation}
which is the similar to the ground-based detectors.
The functions $F^{+,\times}_{a,e}$ are
\begin{equation}
    F^{+}_{a,e}(\lambda_L,\beta_L,\psi_L)= {\rm cos} 2\psi_{L} F^+_{a,e}(\lambda_L,\beta_L) +{\rm sin} 2\psi_{L} F^{\times}_{a,e}(\lambda_L,\beta_L)\;,
\end{equation}
\begin{equation}
    F^{\times}_{a,e}(\lambda_L,\beta_L,\psi_L)= -{\rm sin} 2\psi_{L} F^+_{a,e}(\lambda_L,\beta_L) +{\rm cos} 2\psi_{L} F^{\times}_{a,e}(\lambda_L,\beta_L)\;.
\end{equation}
For $\psi_L=0$, one has
\begin{equation}
    F^{+}_{a}(\lambda_L,\beta_L)= \frac{1}{2}(1+{\rm sin}^2\beta_L){\rm cos}(2\lambda_L-\frac{\pi}{3})\;,
\end{equation}
\begin{equation}
    F^{\times}_{a}(\lambda_L,\beta_L)= {\rm sin}\beta_L {\rm sin}(2\lambda_L-\frac{\pi}{3})\;,
\end{equation}
\begin{equation}
    F^{+}_e(\lambda_L,\beta_L)= \frac{1}{2}(1+{\rm sin}^2\beta_L){\rm cos}(2\lambda_L+\frac{\pi}{6})\;,
\end{equation}
\begin{equation}
    F^{\times}_{e}(\lambda_L,\beta_L)= {\rm sin}\beta_L {\rm sin}(2\lambda_L+\frac{\pi}{6})\;. 
\end{equation}
where the LISA-frame sky position angle $\lambda_L,\beta_L$ and the polarization angle $\psi_L$ are given by 
\begin{equation}
\begin{split}
\lambda_L =\rm arctan \big[\beta \rm cos\lambda(\rm cos\frac{\pi}{3}\rm cos^2\alpha+\rm sin^2\alpha)\\
    +\rm cos\beta \rm sin\lambda \rm cos\alpha \rm sin\alpha(\rm cos\frac{\pi}{3}-1)\\
    +\rm sin\frac{\pi}{3}sin\beta cos\alpha \rm cos\beta \rm sin\lambda\\
    \times(\rm cos\frac{\pi}{3}\rm sin^2\alpha+\rm cos^2\alpha)\\
    +\rm cos\beta \rm cos\lambda \rm cos\alpha \rm sin\alpha(\rm cos\frac{\pi}{3}-1)\\
    +\rm sin\frac{\pi}{3}\rm sin\beta \rm sin\alpha
    \big]\;,
\end{split}
\end{equation}
\begin{equation}
    \beta_L = \rm arsin \big[\rm cos\frac{\pi}{3}\rm sin\beta-\rm sin\frac{\pi}{3}\rm cos\beta \rm cos(\lambda-\alpha)
    \big]\;,
\end{equation}
\begin{equation}
\begin{split}
    \psi_L= \psi + \rm arctan\big[\rm cos\frac{\pi}{3}\rm cos\beta +\rm sin\frac{\pi}{3} \rm sin\beta \rm cos(\lambda-\alpha)\\
    -\rm sin\frac{\pi}{3}\sin(\lambda-\alpha)
    \big]\;,
\end{split}
\end{equation}
with $\alpha=2\pi(t-t_{\rm ref})/1yr$.
$t_{\rm ref}$ is a reference time for the initial position and we assume $t_{\rm ref}=0$.  

The LISA response is both time- and frequency-dependent. 
In the low-frequency limit, because of the motion of the LISA constellation, the time dependency enters both into the Doppler phase term and into the time-dependent LISA-frame angles ($\lambda_L,\beta_L,\Psi_L$) \cite{Marsat:2020rtl}.
Here, we consider the MBHB signals are short, {\it eg}. less than a few days. 
The orientation and position of the LISA constellation in the orbit are barely changed during the GW detection period. 
In the work, we consider the the low-frequency limit and neglect the LISA motion, effectively freeze the LISA constellation in the orbit. We treat the LISA-frame angles ($\lambda_L,\beta_L,\Psi_L$) as constants and neglect both the time and frequency dependency in the response.
In the case, the response is similar to two LIGO-type detectors. The transfer function is just a constant factor.
This approximation can be useful to analytically understand the degeneracies that occurs when using the more complicated full response.
For the short-duration MBHB merger events, it is a reasonable approximation.

\section{Bayesian Methods for massive black hole binary}
\label{sec:inference}

In this section, we will firstly introduce the likelihood construction, and then the dynamic nested sampling methods. 
Bayesian methods are used to estimate the parameters of GW events. In general, the GW data $d(t)$ is a superposition of noise $n(t)$ and a possible signal $h(t)$
\begin{equation}
 d_i(t)=h_i(t)+n_i(t)\;,i=A,E,T
\end{equation}
In our analysis, we will consider the low-frequency limit and simulate the rescaled signals according to Eq. (\ref{tdi}). 

Besides of the waveform in each TDI channels, we also need the noise in each channels. We use the \texttt{tdi} package\footnote{https://lisa-ldc.lal.in2p3.fr} from the LISA Data Challenges Working Group software collection to generate the PSD in each channels, $S^{A,E,T}_{n}$. According to the definition of Eq. (\ref{PSD}), we can obtain the reduced noise PSD, $S^{a,e,t}_{h}$. 
The noise is generated by the public \texttt{pycbc} code \cite{Biwer:2018osg}. For simplicity, one can assume that the noise in the time domain is Gaussian and stationary. Under this assumption, the noise in the frequency domain is Gaussian with zero mean and is characterised by the noise PSD, $\langle \tilde{n}(f)\tilde{n}(f')\rangle=\frac{1}{2}S_n (f)\delta(f-f')$.
Therefore, we can generate the data set by summing the signal and noise in each TDI channels. 

Using the Bayesian methods, the posterior distribution can be determined as
\begin{equation}
    p(\Theta|d,\Lambda)=\frac{p(d|\Theta,\Lambda)p(\Theta|\Lambda)}{p(d|\Lambda)}\;,
\end{equation}
where $\Lambda$ is the model of the GW signal and $\Theta$ are the parameters of this model. $p(\Theta|\lambda)$ represents for the prior probability and the likelihood $p(d|\Theta,\Lambda)$ is the probability of the observed data $d(t)$ given the waveform model $\Lambda$ and a set of parameters $\Theta$.
In the low-frequency limit, the likelihood is given by 
\begin{equation}
    p(d|\Theta,\Lambda) \propto \exp\left[-\frac{1}{2}\langle d-h(\theta)|d-h(\theta)\rangle \right]\;, 
\end{equation}
where the inner product is really the sum of the inner products over all the TDI channels rescaled by Eq. (\ref{tdi})
\begin{equation}
    \langle a|b \rangle =4{\rm Re}\sum_{i=h_a,h_e}\int_0^{\infty}\frac{\tilde{a}^{i}(f)^* \tilde{b}^{i}(f)}{S^{i}_{h}(f)}df\;.
\end{equation}
$S_h^{a,e}(f)$ is noise PSD given by Eq. \ref{PSD}. The optimal signal-to-noise ratio (SNR) is given by 
\begin{equation}
\label{eq:SNR}
    {\rm SNR} = \sqrt{\langle h|h\rangle}\;.
\end{equation}
In this work, we use the reduced TDI templates Eq. (\ref{tdi}) and the analytic function $S^i_{h}$ given by Eq. (\ref{PSD}). The noise PSD for TDI observables $h_a$ and $h_e$ are identical.

Higher-order modes can be used to break the degeneracy between the distance and inclination of the binary coalescence system. With the improvement of the sensitivity of GW detectors, one needs accurate and computationally efficient waveform models, including higher-order harmonics.
In this work, we use the frequency domain waveform model for the inspiral, merger and ringdown of spinning black hole binaries, \texttt{IMRPhenomXHM} \cite{Garcia-Quiros:2020qpx},
which is publicly available as part of the LIGO Algorithm Library Suite (LALSuite) \cite{lalsuite}. The model includes the dominant modes (2,2) and a set of higher-order harmonics,
\begin{equation}
    (l,|m|) =(2,1),(3,3),(3,2),(4,4)\;.
\end{equation}
The model is restricted to the quasi-circular and non-precessing system. 

At last, we use the $\texttt{dynesty}$ package, a public, open-source, python package which implements a dynamic nested sampling methods for inferring Bayesian posteriors distribution of parameters and evidences \cite{Speagle:2019ivv}.
By generating samples in nested "shells", nested sampling is able to estimate evidence as well as the posterior.
And the nested sampling can sample from complex, multi-model distributions.
Compared to Monte-Carlo Markov Chain (MCMC) method, the nested sampling is more suitable to the multi-guassian parameter distribution case. 
Currently, \texttt{pycbc} inference supports the \texttt{dynesty} sample \cite{Biwer:2018osg}.

\section{Validation}
\label{sec:valid}
In this section, we show two test suite results. One example is to show the role of inclusion of the higher-order harmonics in determining the luminosity distance. The other example is to show the capability of testing GR with LISA constellation 

\subsection{Massive black holes signals}
To test the performance of \texttt{LitePIG}, we choose a representative MBHB source. The total redshifted mass $M=m_1+m_2=2.2\times 10^5 M_{\odot}$, mass ratio $q=m_1/m_2=10$ and the redshift $z=1$. We can obtain the luminosity distance from redshift by assuming fiducial cosmology with $H_0=67.1~km\ s^{-1}{\rm Mpc}^{-1}$ and $\Omega_m=1-\Omega_{\Lambda}=0.32$.
We set the inclination angle to $\iota=0.5$ and the dimensionless spin parameter, $a_1=a_2=0$. All the other parameter values are summerized in Table \ref{tab:parameter}. 
Table \ref{tab:snrs} shows the optimal SNR, Eq. (\ref{eq:SNR}), of each harmonic for the representative MBHB merger event.
When calculating the total SNR, the cross terms between modes, $\langle h_{+,\times}^{lm}|h_{+,\times}^{22}\rangle$ can be both positive and negative, causing the interference between the harmonics \cite{Mills:2020thr}.
The (3,2) with the (2,2) mode are the two strongly coupled modes. One of the significant influence of the (3,2) mode can affect the amplitude and phase of the (2,2) mode \cite{Garcia-Quiros:2020qpx}.
Because of this mode mixing effects, the total SNR shown in the "total" column will not be the orthogonal sum of the each mode \cite{2020PhRvD.102b3033K,2014PhRvD..90f4012B,2013PhRvD..87h4004K,London:2020uva}.
\begin{table}[h]
    \centering
    \begin{tabular}{|c|c|}
    \hline
    ${\rm Mass}\ 1 (M_{\odot})$ & $2\times 10^5$\\
    \hline
    ${\rm Mass}\ 2 (M_{\odot})$ & $2\times 10^4$\\
    \hline
    ${\rm Chirp\ mass}(M_\odot)$ &$4.9289\times10^6$\\
    \hline
    $\rm Spin\ 1$ & $0$\\
    \hline
    $\rm Spin\ 2$ & $0$\\
    \hline
    $\rm Redshift$ & $1$\\
    \hline
    $\rm Luminosity\ Distance(Mpc)$ & $6823.1$\\
    \hline
    $\rm Inclination (rad)$ & $0.5$\\
    \hline
    $\rm Ecliptic\ longitude(rad), \lambda$ & $1.5$\\
    \hline
    $\rm Ecliptic\ latitude(rad), \beta$ & $0.57$\\
    \hline
    $\rm Coalescence\ Phase(rad), \phi_0$ & $0.0$\\
    \hline
    $\rm Polarization\ angle(rad), \psi$ & $0.8$\\
    \hline
    ${\rm Coalescence\ time(yr)}, t_0$ & $0.05$\\
    \hline
    \end{tabular}
    \caption{The parameter setup of the simulated MBHB merger. The first column shows the parameters and the units for each parameter. }
    \label{tab:parameter}
\end{table}
\begin{table}[h]
    \centering
    \begin{tabular}{|c|c|c|c|c|c|c|}
    \hline
    ${\rm Modes}$ &${\rm total}$ &$(2,2)$ &$(2,1)$ &$(3,2)$ &$(3,3)$ &$(4,4)$  \\
    \hline
    ${\rm SNR}$     & $447$ &$438$  &$14$    &$6$    &$60$     &$9$\\
    \hline
    \end{tabular}
    \caption{The optimal SNR from each harmonics. Because of the mode mixing, the total SNR is less than the quadrature of all the harmonics.}
    \label{tab:snrs}
\end{table}

In order to generate the waveform of MBHB, we use \texttt{pycbc} package \cite{Biwer:2018osg}, which currently can only generate the waveform of stellar black-hole binary mergers. Hence, we need to rescale the waveform to obtain our targeted waveform from MBHB.
The plus and cross polarization modes of stellar black-hole binary mergers in the frequency domain are assumed to have the following form
\begin{equation}
    \tilde{h}_{+,\times}(f,M_{0})= A(f,M_{0})e^{-i\Psi(f,M_{0})}\;,
\end{equation}
where $M_{0}$ is total mass of stellar black-hole binary.
One needs to rescale the frequency and amplitude. Thus, the GW strains of MBHB with total mass $M$ is given by 
\begin{equation}
    \tilde{h}_{+,\times}(f',M)= A'(f',M)e^{-i\Psi'(f',M)}\;,
\end{equation}
where the frequency is given by
\begin{equation}
f' = f /(M/M_{0})\;,
\end{equation}
the amplitude is obtained from
\begin{equation}
A'(f',M) = A(f, M_{0})\times (M/M_{0})^{2}
\end{equation}
and the phase is idential
\begin{equation}
\Psi(f',M) = \Psi(f,M_{0})\;.
\end{equation}
By rescaling the wavefrom, we can generate plus and cross polarization of MBHB for each harmonic. Fig. \ref{fig:charac_strain} shows the characteristic strain for each harmonic mode.

\begin{figure}
    \centering
    \includegraphics[width=1.0\linewidth]{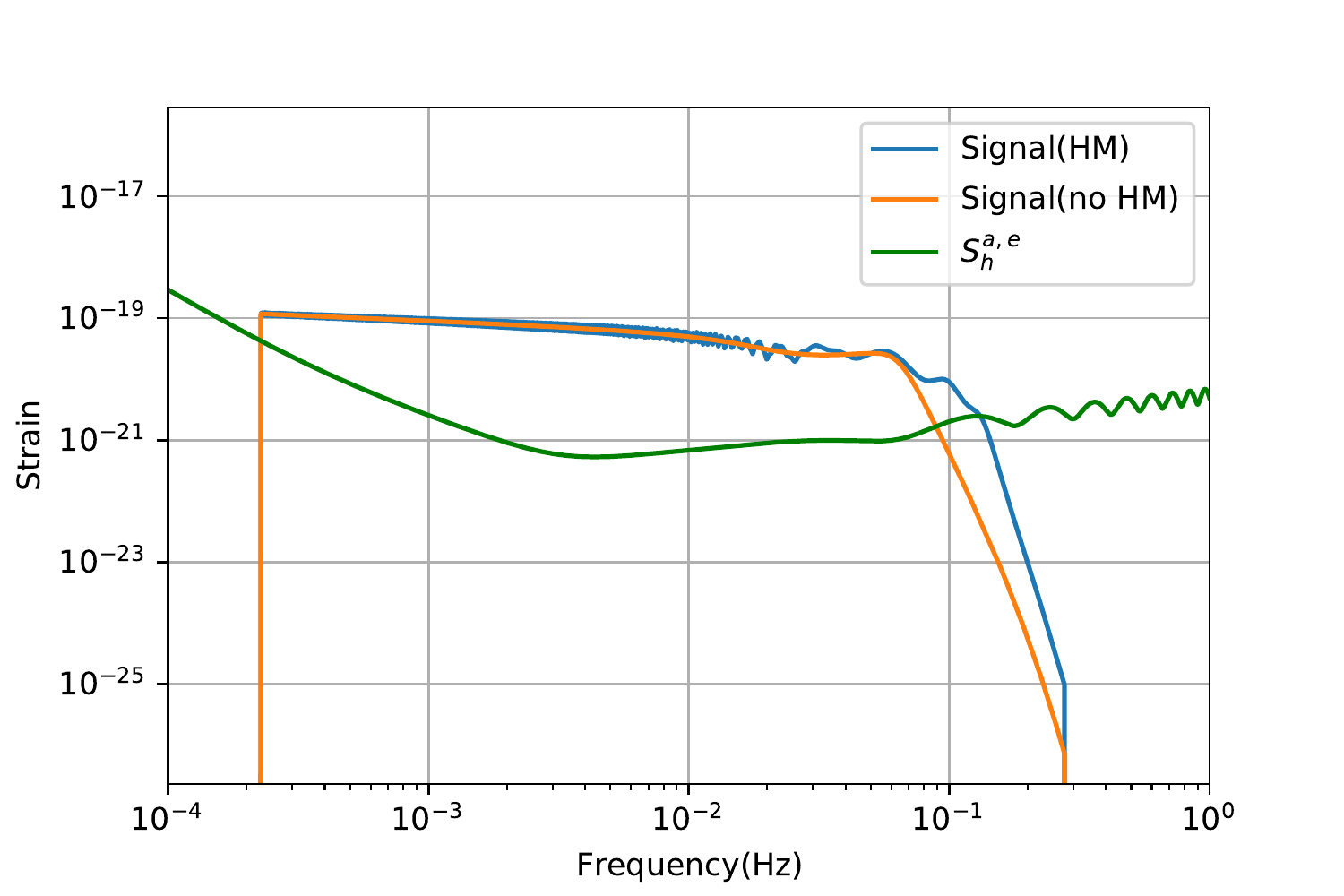}
    \caption{Characteristic strain of full signal compared to the (2,2) dominant mode. The blue curve is the strain of the full signal which is the sum of the individual mode. The orange curve show the strain of the (2,2) dominant mode. The green solid curves show the characteristics strain with the reduced noise PSD $S^{a,e}_h(f)$ from Eq. (\ref{PSD})}
    \label{fig:signal}
\end{figure}

Fig. \ref{fig:signal} shows the response for the TDI observables Eq. (\ref{LFtdi}) in the cases with and without higher-order modes.
Though the individual harmonics have fairly smooth amplitude shown in Fig.(\ref{fig:charac_strain}), the full signal shows obvious wiggles. This is because the frequency dependence of different harmonics are different. 

\subsection{Parameter estimation}

\begin{figure}[h]
    \centering
    \includegraphics[width=1.0\linewidth]{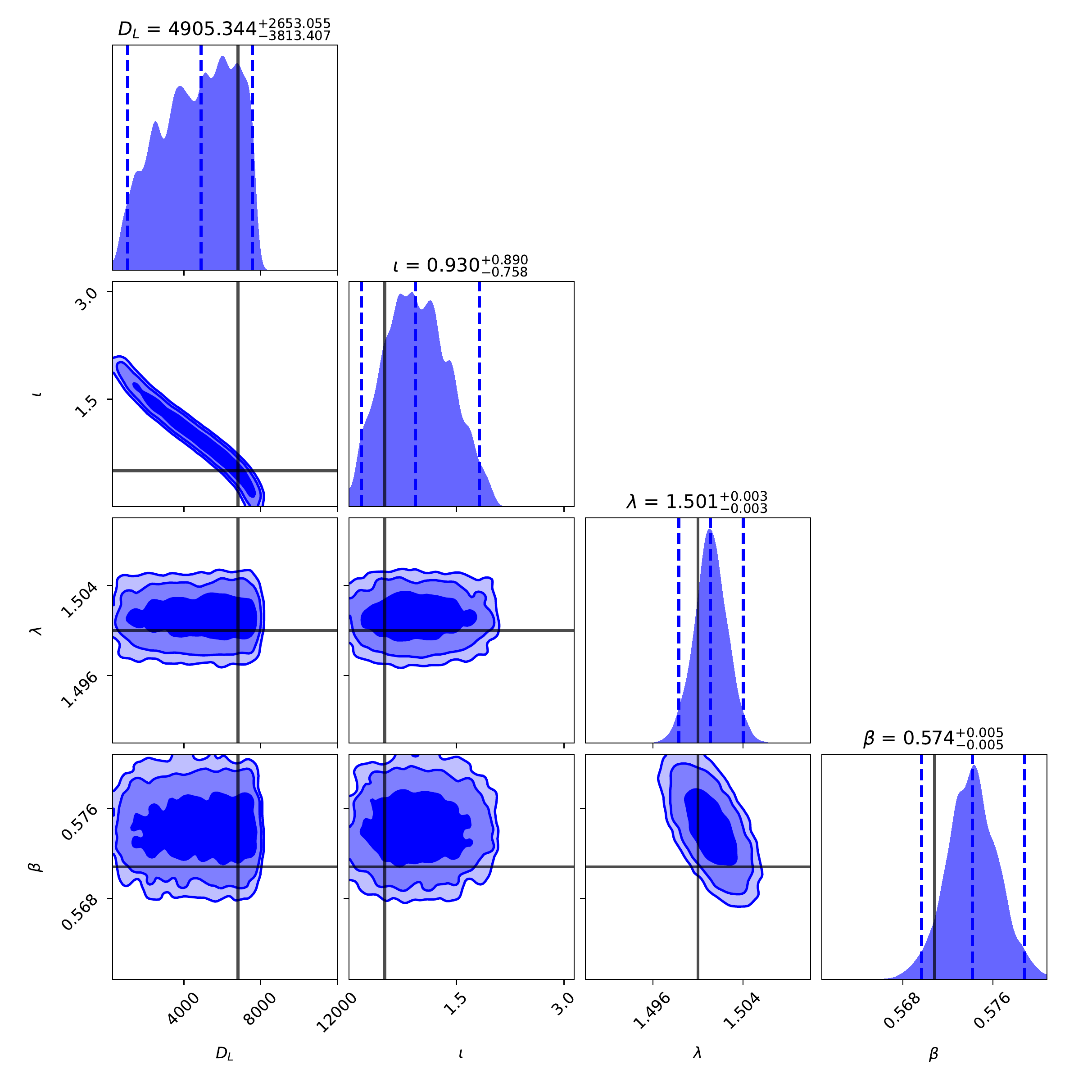}
    \caption{The posterior distribution on the parameters of the MBHB without the higher-order modes. The vertical black lines are the true injection values. 
    The vertical dashed lines show the $2\sigma$ credible interval. 
    The posterior distributions show the $1\sigma$,$2\sigma$ and $3\sigma$ contours.}
    \label{fig:posterior_noHM}
\end{figure}

\begin{figure}[h]
    \centering
    \includegraphics[width=1.0\linewidth]{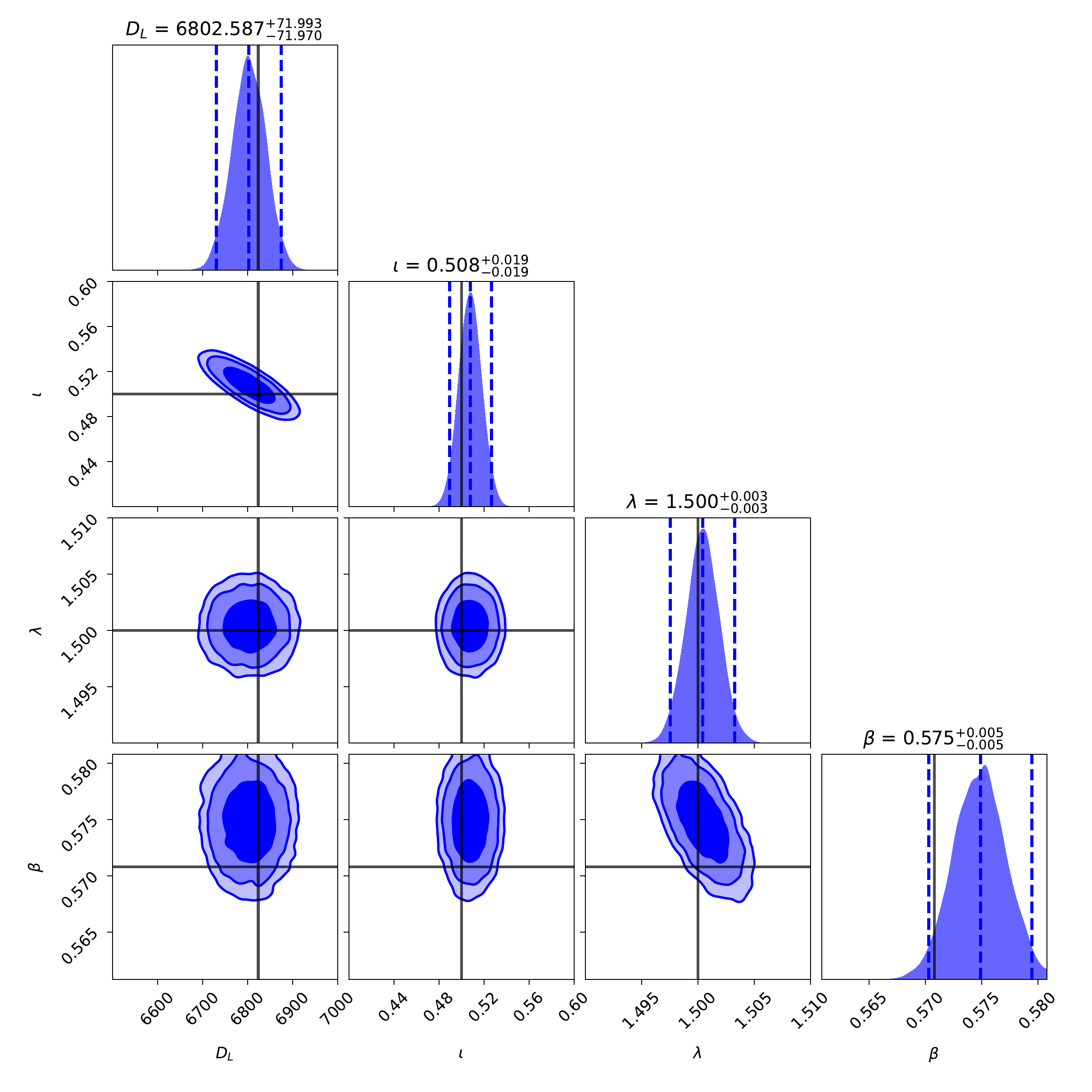}
    \caption{The posterior distribution on the parameters of the MBHB with the higher-order modes. The vertical black lines are the true injection values. The posterior distributions show the $1\sigma$,$2\sigma$ and $3\sigma$ contours}
    \label{fig:posterior_HM}
\end{figure}

For the momentum, we do not perform the initial search for the alerted events.  Instead, we assume a source has already been identified in the data stream.
The priors are give in Table \ref{tab:prior}. The parameter priors for $\{q,a_1,a_2,D_L,\lambda,\psi,\phi_0\}$ are uniform. The log-uniform prior is used on chirp mass. The $\iota$ prior is uniform in ${\rm cos}\iota$ and the $\beta$ prior is uniform in ${\rm sin} \beta$.
\begin{table}[h]
    \centering
    \begin{tabular}{|c|c|c|}
    \hline
      ${\rm Parameter}$  &$ {\rm Lower~Bound} $& ${\rm Upper~Bound}$ \\
      \hline
      $\log M_c$      &$\log(10^4)$    &$\log(10^7)$\\
      \hline
      $q$          &$1$           &$10^3$\\
      \hline
      $a_1$        &$-1$          &$-1$\\
      \hline
      $a_2$        &$-1$          &$-1$\\
      \hline
      $D_L$        &$10^{-2}$     &$10^4$\\
      \hline
      $\lambda$    &$0$           &$2\pi$\\
      \hline
      $\sin \beta$  &$-1$          &$1$\\
      \hline
      $\cos \iota$ &$-1$          &$1$\\
      \hline
      $\psi$       &$0$           &$\pi$\\
      \hline
      $\phi_0$      &$0$           &$2\pi$\\
      \hline
      $t_0$        &$0$           &$1$\\
      \hline
    \end{tabular}
    \caption{The lower and upper bounds of the the prior distributions in our Bayesian inference. We assume all priors are uniformly distributed except for the prior on ($M_c,\iota,\beta$). We assume uniform distributions for isotropically distributed angles on the sphere, so the prior on the inclination angle and ecliptic latitude are uniform distribution on ${\rm cos}\iota$ and ${\rm sin}\beta$.}
    \label{tab:prior}
\end{table}
We adopt the dynamic nested sampler.
In the following analysis, we will present the corner plots of the posterior distribution of the parameters.
Fig. \ref{fig:posterior_noHM} and Fig. \ref{fig:posterior_HM} show the posterior distribution on the parameters in the cases without and with the higher-order harmonics. 
The mean and standard deviation can be found in Table \ref{tab:error}.
In the case without the higher-order modes, there are apparent bias for the parameters in the posterior distribution.
The higher-order harmonics can reduce the estimation parameter errors on luminosity distance and inclination angle roughly by a factor of 50-ish. 
This is because the higher-order modes can break the degeneracy between distance and inclination angle. 

\begin{table}[h]
    \centering
    \begin{tabular}{|c|c|c|c|}
    \hline
    ${\rm Parameter}$ &${\rm Injection}$ &${\rm w/o~HM}$ &${\rm w/~HM}$  \\
    \hline
    $D_L$        &$6823.1$ &$4905.3_{-2376.7}^{+1890.4}$ &$6802.6_{-35.0}^{+36.8}$      \\
    \hline
    $\lambda$    &$1.5$  &$1.501_{-0.001}^{+0.001}$ &$1.500_{-0.001}^{+0.001}$          \\
    \hline
    $\beta$  &$0.57$  &$0.574_{-0.002}^{+0.002}$  &$0.575_{-0.002}^{+0.002}$         \\
    \hline
    $\iota$ &$0.50$  &$0.930_{-0.439}^{+0.494}$ &$0.508_{-0.010}^{+0.009}$      \\
    \hline
    \end{tabular}
    \caption{The $1\sigma$ errors for each parameters.}
    \label{tab:error}
\end{table}

\subsection{Test GR}
The existence of the higher-order modes opens a new window for testing GR.
This is because higher-order harmonics carry fruitful information in the ringdown phase of CBC, which corresponds to the strong gravitational field regime.
Hence, the test whether the amplitude of higher-order harmonics are consistent with the predictions of GR, deliver a completely new message about the gravitational dynamics close to the black hole horizon.
In general, the GW tests of GR can be divided into two categories. The first category is to test the phase evolution. In the second test, one looks for anomalies in the amplitudes of the higher-order modes \cite{Islam:2019dmk}.
In our work,  we follow \cite{Islam:2019dmk} and allow for deviation in the amplitudes of the higher-order modes
\begin{equation}
\begin{split}
    h_+ -i h_{\times}=
    \sum_{m=\pm2}\ _{-2}Y_{2m}(\iota,\phi_0)h_{2m}\\
    +\sum_{l>2}\sum^{l}_{m=-l}(1+c_{lm})\ _{-2}Y_{lm}(\iota,\phi_0)h_{lm}\;,
\end{split}
\end{equation}
where the $c_{lm}$ are the free parameters. In the case of GR, $c_{lm}=0$.
Since $h_{l,-m}=(-1)^lh_{lm}^*$, we have $c_{lm} =c_{l,-m}$.
We use the \texttt{IMRPhenomXHM} template and focus on all mode $(l,|m|) =(2,2),(2,1),(3,3),(3,2),(4,4)$.
We simulate the signals generated by MBHB merger and noise according to LISA setup . 
The source parameter values are kept as the same in Table \ref{tab:parameter}. 
The only difference is we allow all the $c_{lm}$ can vary simultaneously. 
We assume GR as the fiducial model, so we choose values for the deviation parameter $c_{lm}=0$ in the injection.
\begin{table}[h]
    \centering
    \begin{tabular}{|c|c|c|c|}
    \hline
        $c_{21}$ & $c_{32}$ & $c_{33}$ & $c_{44}$\\
    \hline
    $-0.65^{+0.22}_{-0.08}$ & $-0.65^{+0.22}_{-0.08}$ & $0.56^{+0.60}_{-0.76}$ & $1.57^{+2.34}_{-1.90}$\\
    \hline
    \end{tabular}
    \caption{1$\sigma$ deviation from GR of the higher-order harmonic coefficients.}
    \label{tab:clm}
\end{table}

\begin{figure}
    \centering
    \includegraphics[width=1.0\linewidth]{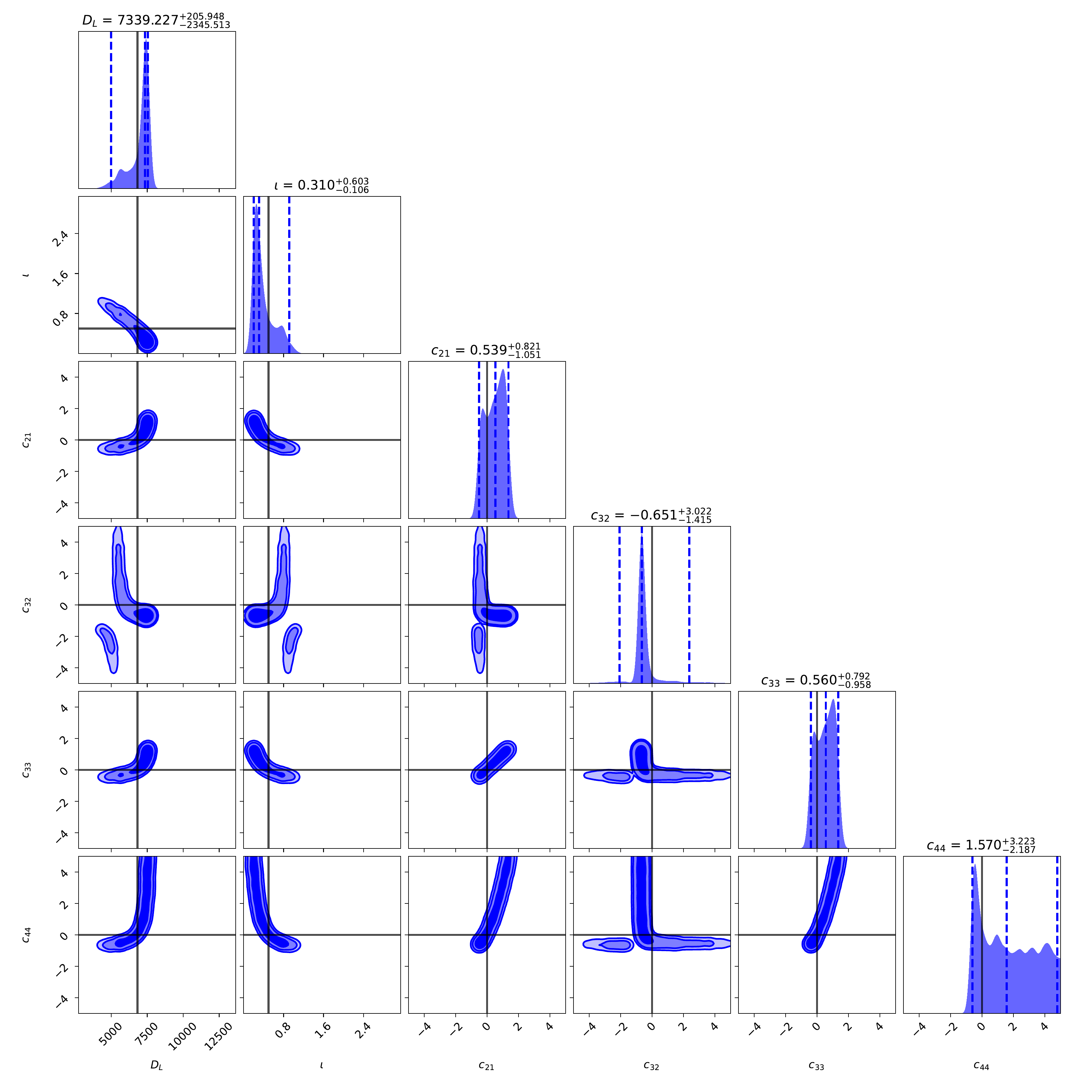}
    \caption{The posterior distribution on $c_{lm}$ for MBMB merger. The vertical black lines are the true injection value. The vertical blue lines indicate $95\%$ confidence intervals.}
    \label{fig:test_GR}
\end{figure}

Fig. \ref{fig:test_GR} show the posterior probability distribution in the GR test. Because the mode amplitudes can vary independently, the inclusion of the higher-order harmonics will not significantly help to break the inclination-distance degeneracy.  Hence, one can see that the luminosity distance error in the GR test is much larger than those assuming GR. As for the test of GR,  we list the results in Table \ref{tab:clm}. Because of the limitation of data quality, the current GW data detected by LVK can barely give a strict constraint on these parameters \cite{Puecher:2022sfm,LIGOScientific:2016lio,LIGOScientific:2018dkp,LIGOScientific:2019fpa,LIGOScientific:2020tif,LIGOScientific:2021sio}. Take GW190412 as an example \cite{Puecher:2022sfm}, the posterior on $c_{33}$ is bimodal distributed, due to the degeneracy with the inclination angle. The corresponding errors are almost an order of magnitude worse than here we got. Furthermore, these parameter constraint results are obtained by varying these $c_{lm}$ one-by-one, not simultaneously as here we did.  

\section{Conclusions}
In this paper, we present a new parameter inference system for the gravitational wave data in the millihertz band. We are aiming for the compact binary coalescence originated from the massive black hole binary. In particular, we highlight the role of higher-order harmonics. We show that by including the first four higher-order modes, the famous distance-inclination degeneracy can be very effectively broken. The corresponding errors on the luminosity distance and inclination angle can be reduced roughly by a factor of 50-ish. Furthermore, we show the capability of testing general relativity by LISA constellation. The superb sensitivity of LISA and the robust GW signals from MBHB allow us to detect the ringdown phase of compact binary coalescence process with fairly high signal-to-noise ratio. Hence, it opens a new window to explore the strong gravitational field regime, which is the scale of a few times the black hole event horizon. Our results show that the GW from MBHB can simultaneously constrain four of the higher harmonic amplitudes (deviation from GR) with a precision of $c_{21}=0.54^{+0.61}_{-0.82}$, $c_{32}=-0.65^{+0.22}_{-0.08}$, $c_{33}=0.56^{+0.60}_{-0.76}$ and $c_{44}=1.57^{+2.34}_{-1.90}$, respectively.  

The software package incorporate the following features: the GW waveform emitted from the massive black hole binaries, the stationary instrumental gaussian noise, the higher-order harmonic modes, the full response function from the time delay interferometry (TDI) and the gaussian likelihood function with the dynamic nested parameter sampler. The resulting code, \texttt{LitePIG}, is based on the widely used waveform generator \texttt{pycbc}, which is currently only suitable for the GW emitted from the stellar-mass black hole. We extended this waveform to the MBHB case by some rescaling law. The LISA instrumental noise is imported from the LISA data challenge (\texttt{LDC}) working package. All of the libraries, which \texttt{LitePIG} is based on, are the mature tools that are widely adopted by the GW community. Hence, we believe \texttt{LitePIG} is a reliable package for the GW parameter estimation.
The code is publicly available on the repository  \url{https://github.com/renjiewang888/LitePIG.git}.

\section*{Acknowledgements}
We thank Zhoujian Cao, Xiaolin Liu and Junjie Zhao for the discussion on the gravitational waveform generation. This work is supported by the National Key R\&D Program of China No. 2021YFC2203001.



\bibliography{HM}

\end{document}